\documentclass{PoS}

\newcommand{\SU}{\textrm{SU}}
\newcommand{\U}{\textrm{U}}
\newcommand{\SO}{\textrm{SO}}
\newcommand{\comma}{\ ,}
\newcommand{\period}{\ .}
\newcommand{\tr}{\textrm{tr}}

\renewcommand{\Re}{\textrm{Re}}

\title{Fermions in higher representations. Some results about SU(2) with adjoint
fermions. }

\ShortTitle{Fermions in higher representations. Some results about SU(2) with
adjoint fermions. }

\author{Luigi Del Debbio \\ 
        SUPA, School of Physics and Astronomy, University of Edinburgh,
        Edinburgh EH9 3JZ, Scotland\\
        E-mail: \email{luigi.del.debbio@ed.ac.uk} 
} 
\author{\speaker{Agostino Patella} \\
        Department of Physics, Swansea University,
        Singleton Park, Swansea SA2 8PP, Wales\\
        E-mail: \email{a.patella@swansea.ac.uk}
}
\author{Claudio Pica \\
        Physics Department, Brookhaven National Laboratory,
        Upton, NY 11973-5000, USA\\
        E-mail: \email{pica@quark.phy.bnl.gov} 
} 

\abstract{
We discuss the lattice formulation of gauge theories with fermions in arbitrary
representations of the color group, and present the implementation of the RHMC
algorithm for simulating dynamical Wilson fermions. A first dataset is presented
for the $\SU(2)$ gauge theory with two fermions in the adjoint representation,
which has been proposed as a possible technicolor candidate. Simulations are
performed on $8^3 \times 16$ lattices, at fixed lattice spacing. The PCAC mass,
the pseudoscalar, vector and axial meson masses, the pseudoscalar meson decay
constant are computed. The extrapolation to the chiral limit is discussed.
However more extensive investigations are needed in order to control the
systematic errors in the numerical results, and then understand in detail the
phase structure of these theories.
}

\FullConference{The XXVI International Symposium on Lattice Field Theory \\
                 July 14 - 19, 2008\\
                 Williamsburg, Virginia, USA}

\begin{document}

\section{Introduction}

Gauge theories with fermions in two-index representations play important roles
in different scenarios for Physics beyond the Standard Model (SM):
supersymmetric theories, technicolor
models~\cite{Weinberg:1975gm,Susskind:1978ms,Hill:2002ap,Sannino:2008ha},
orientifold theories~\cite{corrigan:1979xf,Armoni:2003gp}. So far the lattice is
the unique tool for studying strongly-interacting theories from first
principles. Moreover recent algorithmic progress and the increase in computer
performance allow these theories to be studied by Monte Carlo simulations at an
affordable cost. On the other hand the expertise developed in many years of
simulating lattice QCD must be exported with some care as we try to interpret
results for generic strongly-interacting theories, since these may reveal a very
different phenomenology from QCD.

Therefore the challenge this work attempts to address is twofold. We generalize
the standard RHMC algorithm for simulating QCD to the case of $\SU(N)$ gauge
group and fermions in a generic representation. We briefly describe the main
characteristics of our bespoke code -- HiRep -- in which we have implemented the
two-index representations for generic number of colors. Then we present some
preliminary result for $\SU(2)$ with two Dirac fermions in the adjoint
representation.

The number of fermions in this theory is close to the estimated critical value
where an infrared (IR) fixed point is expected to appear~\cite{Banks:1981nn}. If
an IR fixed point exists, the theory is conformal in the IR, in particular it
does not confine and does not break chiral symmetry. If the theory does not have
the IR fixed point, the vicinity to the critical number of flavours can induce a
walking behaviour of the renormalized coupling constant, which would make such a
theory a possible realization of the technicolor
mechanism~\cite{Dietrich:2006cm,Foadi:2007ue}. Which scenario is
the correct one can only be established by numerical simulations on the lattice.
However for realistic numerical simulations, the finite volume, the non-zero
fermion mass, and the lattice artifacts introduce systematic errors that are
likely to obscure the behaviour one would naively infer from the existence of an
IR fixed point. In order to obtain clear and robust results, we need to go to
\emph{large} volumes, \emph{small} masses and \emph{small} lattice spacing,
keeping in mind that we do not know a priori what \emph{large} and \emph{small}
mean in this case.

Presenting the results, we will try to emphasize which are compatible with a
standard QCD-like scenario, and which deviate from it giving hints of new
interesting dynamics.

At the end of this introduction, we want to point out that the same theory was
explored in~\cite{Catterall:2007yx}; other gauge theories close to the conformal
window have recently been investigated
in~\cite{Appelquist:2007hu,Shamir:2008pb,Deuzeman:2008sc} (see also the talks of
De Grand, Deuzeman, Fleming, Hietanen, Holland, Jin, Neil, Nogradi, Pallante,
Svetitsky in this conference).

\section{Description of the HiRep code}

The HiRep code implements the RHMC algorithm~\cite{Clark:2006fx} for simulating
dynamical Wilson fermions in a generic representation of $\SU(N)$, once the
mathematical definition of the representation is provided. One and two-index
representations are implemented. In what follows, we summarize the main features
of the implementation (for details, see~\cite{deldebbio:2008zf}).

\begin{description}

\item[Generic representations.] We call $T^a$ the generators of the $\SU(N)$
group, $T^a_R$ the represented generators in the representation $R$, and $T_R$
their normalization defined as $\tr [ T^a_R T^b_R ] = T_R \delta_{ab}$. For each
link, we store both the link-variable in the fundamental representation
$U(x,\mu)$ and the represented one, defined by:
\begin{equation}
U_R(x,\mu) = \exp \left[ i \omega^a(x,\mu) T^a_R \right] \comma
\quad \textrm{with} \quad
U(x,\mu) = \exp \left[ i \omega^a(x,\mu) T^a \right] \period
\end{equation}
In the code, the generators $T^a$, $T^a_R$ an the definition of $U_R$ are
implemented for one and two-index representations.

\item[Dirac operator.] The Wilson discretization of the Dirac operator is
obtained from the naive one by simply replacing the link-variables with the
represented ones:
\begin{equation}
D \psi(x) = \psi(x) - \kappa \sum_\mu \left\{
( 1 - \gamma_\mu) U_R(x,\mu) \psi(x + \hat{\mu}) + 
( 1 + \gamma_\mu) U_R(x-\hat{\mu},\mu)^\dagger \psi(x - \hat{\mu})
\right\} \period
\end{equation}
We also define the Hermitian operator $Q = \gamma_5 D$.

\item[MD Hamiltonian.] The determinant of the Dirac operator in the partition
function is stochastically estimated:
\begin{equation}
\det D^{n_f} = \det Q^{n_f} \propto
\int \exp \left\{ - \phi^\dagger Q^{-n_f} \phi \right\}
\mathcal{D}\phi \mathcal{D} \phi^\dagger \comma
\end{equation}
and the generic negative powers of $Q^2$ are computed by means of rational
approximations $x^{-p/2} \simeq \sum_i \alpha^{(p)}_i / (x - \beta^{(p)}_i)$,
\begin{equation}
Q^{-n_f} = (Q^2)^{-n_f/2} \simeq
\sum_i \alpha^{(n_f)}_i \left[ Q^2 - \beta^{(n_f)}_i \right]^{-1} \period
\end{equation}
The inversions in the expressions above are performed simultaneously thanks to
the implementation of a multi-shift inverter.
The momenta $\pi(x,\mu)$ of the molecular dynamics (MD) are introduced, as
elements of the Lie algebra, $\pi(x,\mu) = i \pi^a(x,\mu) T^a$. The MD
Hamiltonian is finally written as $H = H_\pi + H_g + H_f$ with:
\begin{eqnarray}
&& H_\pi = \frac12 \sum_{x,\mu} \tr [\pi(x,\mu)^2] =
\frac{T_F}{2} \sum_{x,\mu,a} \pi^a(x,\mu)^2 \comma \\
&& H_g = -\frac{\beta}{2N} \sum_{x, \mu \neq \nu} \tr \mathcal{P}_{\mu \nu}(x) \comma \\
&& H_f = \sum_{x,i} \alpha^{n_f}_i
\phi^\dagger (x) \left[ Q^2 - \beta^{(n_f)}_i \right]^{-1} \phi(x) \period
\end{eqnarray}
Also multiple pseudofermion actions and even-odd preconditioning are
implemented; for details see~\cite{deldebbio:2008zf}.

\item[MD evolution equations.] Denoting with $\tau$ the MD time, the evolution
equations for the MD Hamiltonian are:
\begin{equation}
\frac{d}{d\tau} U(x,\mu) = \pi(x,\mu) U(x,\mu) \comma
\qquad
\frac{d}{d\tau} \pi^a(x,\mu) = F^a_g(x,\mu) + F^a_f(x,\mu) \comma
\label{eq:md_ev}
\end{equation}
where the $F_g$ is the usual gauge force, while $F_f$ is the usual fermions
force where all the link-variables and the generators are replaced with the
corresponding quantities in the representation $R$ (see~\cite{deldebbio:2008zf}
for explicit computation):
\begin{eqnarray}
&& F^a_g(x,\mu) = \frac{\beta}{T_F N} \Re \ \tr
\left[ i T^a U(x,\mu)V(x,\mu)^\dagger \right] \comma \\
&& F^a_f(x,\mu) = - \frac{1}{T_F} \Re \ \tr
\Big\{ i T^a_R U_R(x,\mu) \gamma_5 ( 1 - \gamma_\mu ) \times
\nonumber \\
&& \qquad \qquad \qquad \times
\sum_i \alpha^{(n_f)}_i \left[
\eta_i(x+\hat{\mu}) \xi_i(x)^\dagger + \xi_i(x+\hat{\mu}) \eta_i(x)^\dagger
\right] \Big\} \comma
\end{eqnarray}
where $V(x,\mu)$ is the staple in the link $(x,\mu)$ and the pseudofermions
$\eta_i$ and $\xi_i$ are defined as $\eta_i = \left[ Q^2 - \beta^{(n_f)}_i \right]^{-1} \phi$, and $\xi_i = Q \eta_i$.
The Eqs.~(\ref{eq:md_ev}) are integrated using a
second order Omelyan integrator~\cite{Takaishi:2005tz}; moreover the
exponentiation of the momenta is approximated with an exact unitary
matrix~\cite{Luscher:2005rx}.

\item[Random updates.] Before each MD trajectory, the momenta and the
pseudofermions must be chosen with the random distributions defined by the
partition function. In particular each component of the momenta is independently
generated with Gaussian distribution proportional to $\exp \left( - T_F \pi^a(x,\mu)^2 /2 \right)$. While $\phi$ is obtained by generating a pseudofermion $\tilde{\phi}$ with
distribution proportional to $\exp ( - \tilde{\phi}^\dagger \tilde{\phi} )$, and defining:
\begin{equation}
\phi = Q^{-n_f/2} \tilde{\phi} = (Q^2)^{-n_f/4} \tilde{\phi} \simeq
\sum_i \alpha^{(n_f/2)}_i \left[ Q^2 - \beta^{(n_f/2)}_i \right] \tilde{\phi}
\period
\end{equation}

\item[Metropolis test.] In order to correct the error introduced by the numeric
integration of the MD evolution equations, the new configuration generated by
the MD evolution is accepted/rejected with a standard Metropolis test.

\end{description}

\section{The lattice simulations}

We simulated $\SU(2)$ gauge theory with two Dirac fermions in the adjoint representation, on $16 \times 8^3$ lattices, at fixed $\beta=2.0$ and various $\kappa$.

\begin{figure}[ht]
\centering
\includegraphics[width=0.45\textwidth]{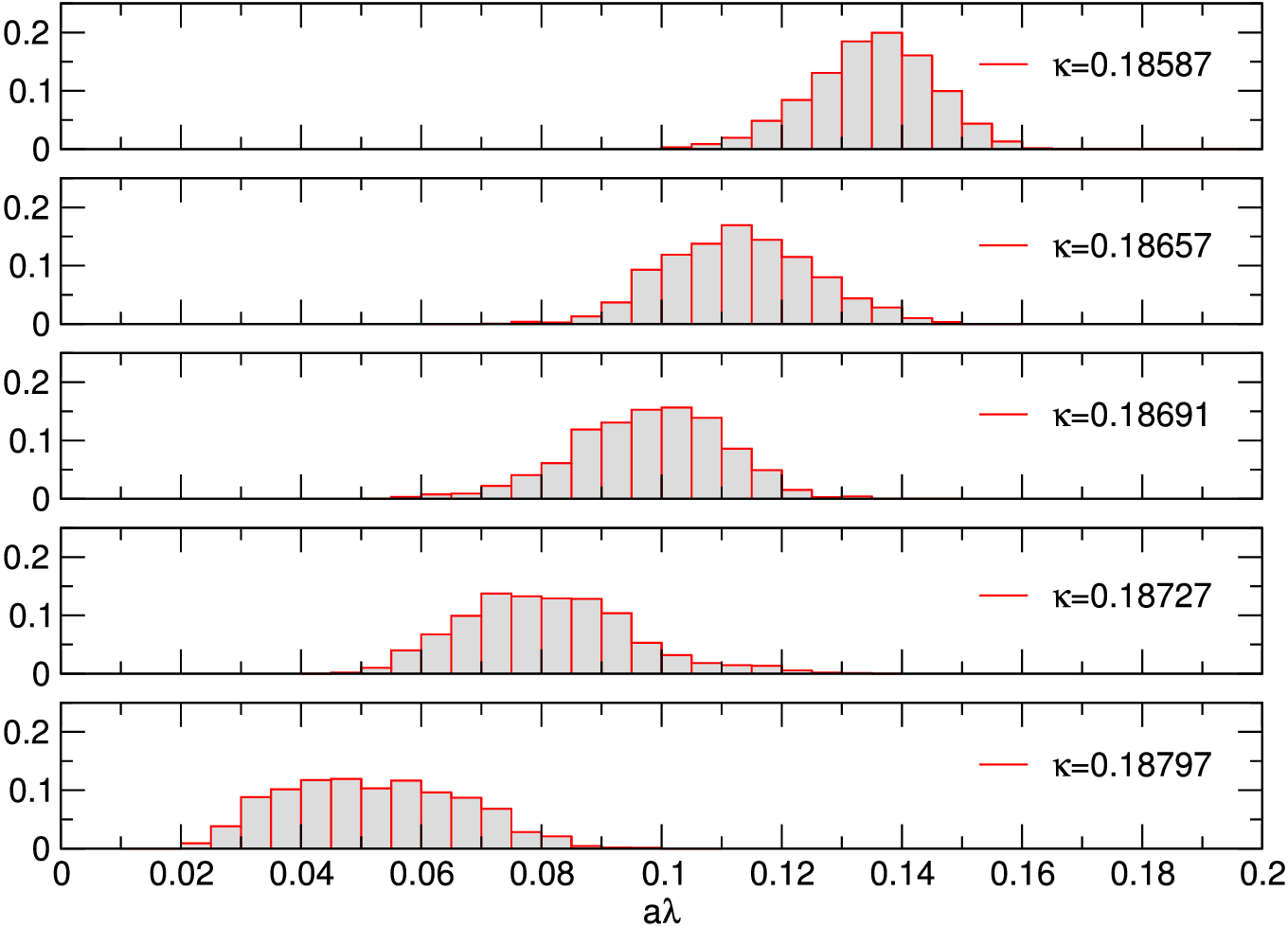}
\label{fig:lowev}
\caption{Probability distributions of the lowest eigenvalue of $Q^2$. }
\end{figure}

In order to check that the Dirac operator has no exceptionally small
eigenvalues, we have monitored the distribution of the lowest eigenvalue of
$Q^2$ (Fig.~\ref{fig:lowev}). The results obtained with fundamental fermions
in~\cite{DelDebbio:2005qa} suggest that the width of the distribution is
proportional to $a/\sqrt{V}$, therefore the existence of exceptionally small
eigenvalues is suppressed in the limit of infinite volume or zero lattice
spacing (at fixed value of $\kappa$). On the other hand at fixed volume and
lattice spacing the probability of getting exceptionally small eigenvalues is
non-zero if we go close enough to the chiral limit. This region of parameter
space is also likely to be affected by the presence of an Aoki phase around the
chiral point, which is expected to break the flavour symmetry with pattern
$\SO(4) \rightarrow \U(1) \times \U(1)$ in the theory under
scrutiny~\cite{DelDebbio:2008wb}. Also the Aoki phase is a lattice artifact and
it is expected to vanish in the continuum limit. It is clear that even if we are
observing neither exceptionally small eigenvalues nor signatures of the Aoki
phase, if we are too close to the chiral limit our results may be affected by
large systematic error. The only way to get reliable results is to check their
stability while increasing the volume and reducing the lattice spacing. This has
not been done in this work, in which we present data at fixed volume and fixed
lattice spacing, and it remains one of our main goals for the future.

We measured the PCAC mass $m$ of the quarks, the pseudoscalar meson mass
$M_{PS}$, its decay constant $F_{PS}$, the vectorial meson mass $M_V$, and the
axial meson mass $M_A$. The results are listed in Table~\ref{tab:pheno}. The
masses are extracted by fitting the plateau of the effective mass extracted from
the correlator. On the lattices considered so far this plateau
consists only of few points, and this is another source of systematic error.

\begin{table}[ht]
\centering
\begin{tabular}{ccccccc}
$\kappa$ & $a m$ & $aM_{PS}$ & $aM_{PS}^2/m$ & $aF_{PS}$ & $aM_V$ & $M_A/M_V$ \\
\hline
0.18587 & 0.2209(30) & 1.149(11) & 5.98(14) & 0.399(14) & 1.269(12) & 1.51(22) \\
0.18657 & 0.1874(40) & 1.044(18) & 5.82(20) & 0.354(20) & 1.163(20) & 1.75(26) \\
0.18692 & 0.1559(22) & 0.919(13) & 5.41(15) & 0.300(11) & 1.029(14) & 1.45(7) \\
0.18727 & 0.1236(25) & 0.764(20) & 4.72(25) & 0.247(12) & 0.850(25) & 1.10(10) \\
0.18748 & 0.1012(20) & 0.652(21) & 4.20(27) & 0.215(10) & 0.733(22) & 1.00(10) \\
0.18769 & 0.0720(17) & 0.485(28) & 3.26(37) & 0.184(7)  & 0.511(34) & 0.75(12) \\
0.18797 & 0.0238(8)  & 0.278(25) & 3.26(59) & 0.102(8)  & 0.275(32) & 0.69(17) \\
\hline
\end{tabular}
\caption{Fitted values for the masses and decay constant in lattice units, in
various interesting combinations.}
\label{tab:pheno}
\end{table}

\begin{figure}[!ht]
\centering
\includegraphics[width=0.75\textwidth]{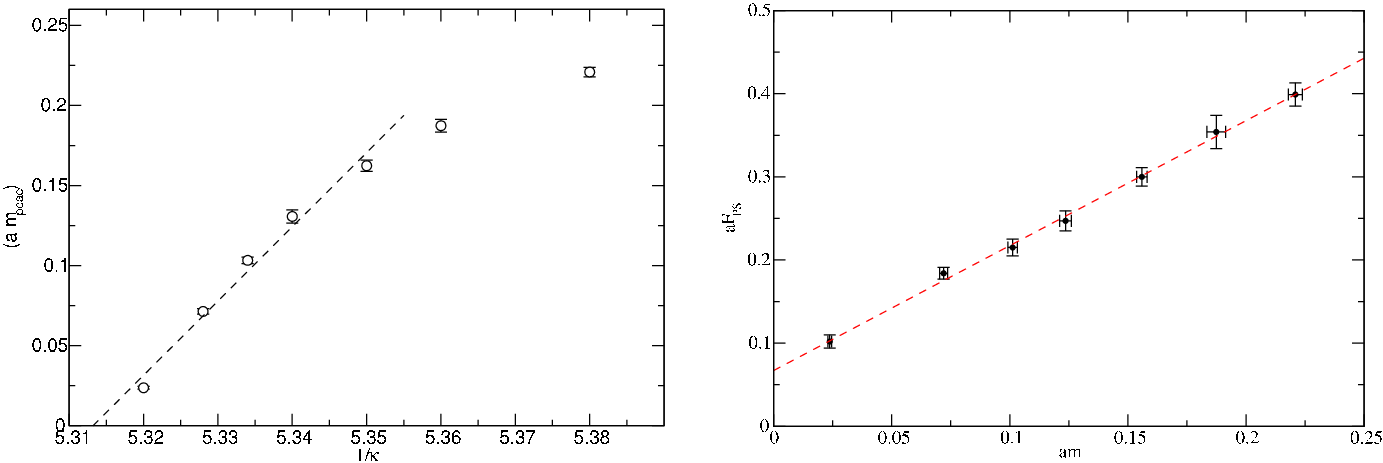}
\label{fig:kc_fps}
\caption{On the left side, linear extrapolation of the PCAC mass as a function of the inverse
hooping parameter $1/\kappa$. Only the five lowest points are used for the
fit. On the right side, linear extrapolation of the PS decay constant as a function of the PCAC
mass.}
\end{figure}

The chiral point can be identified by fitting the PCAC mass with a linear
function of $1/\kappa$. This fit is reasonable (see Fig.~\ref{fig:kc_fps}) and
yields $\kappa_c = 0.18827(4)$.

If the standard scenario of the chiral symmetry breaking were realized, then the
ratio $a M_{PS}^2/m$ should become a constant in the chiral limit. Although it goes
flat in the two lightest points (see Table~\ref{tab:pheno}), it has an unexpected
behaviour, dropping rapidly from the heaviest points and then stopping abruptly.
Robust conclusions can be drawn in this case and a more careful study of
the chiral limit is required.

On the contrary, the linear behaviour of the PS decay constant with the PCAC
mass is pretty clear. A linear extrapolation can be made using all the points
and the result for the chiral limit is $aF_{PS}(m=0) = 0.084(4)$
(Fig.~\ref{fig:kc_fps}).

\begin{figure}[!ht]
\centering
\includegraphics[width=0.78\textwidth]{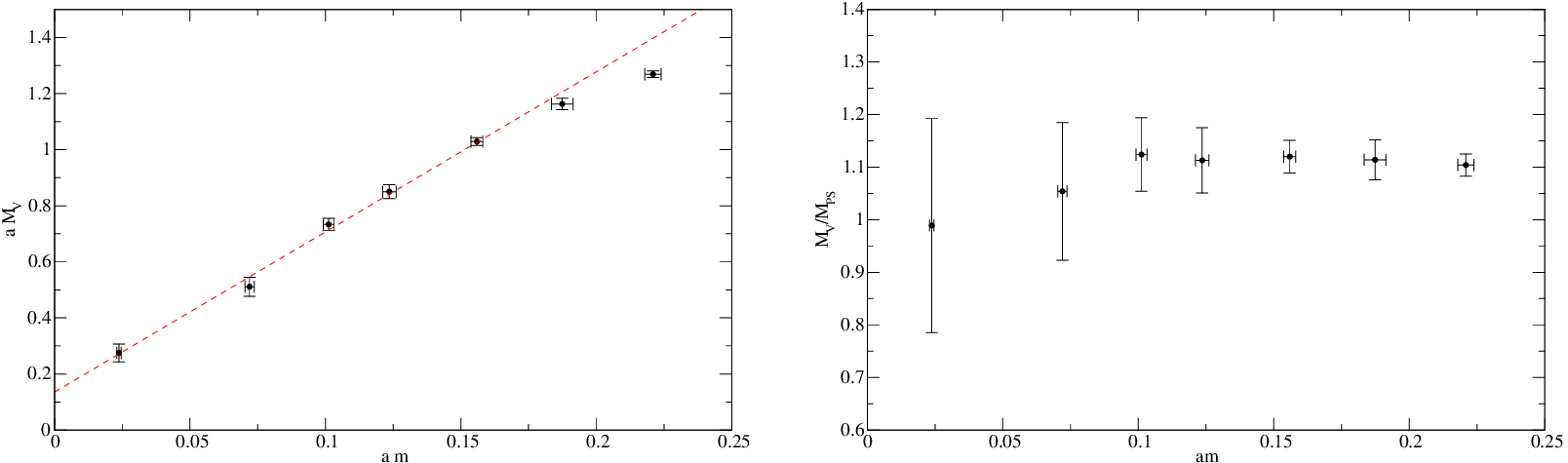}
\label{fig:mv}
\caption{On the left, linear extrapolation of the vectorial meson mass as a
function of the PCAC mass. On the right, the ratio $M_V/M_{PS}$ as a
function of the PCAC mass.}
\end{figure}

The linear behaviour of the V mass with the PCAC mass is quite good but it is
not clear if the vectorial meson stays massive in the chiral limit
(Fig.~\ref{fig:mv}). The ratio $M_V/M_{PS}$ is also plotted in
Fig.~\ref{fig:mv}; in the chiral symmetry were spontaneously broken, this ratio
should diverge in the chiral limit, while we observe here that it stays finite
(and compatible with $1$).

Finally the ratio $M_A/M_V$ is reported in the last column of
Table~\ref{tab:pheno}. The lightest points seems to indicate an inversion of the
vectorial and axial mesons, and therefore a violation of the second Weinberg sum
rule, which is another hint for non-standard physics.

\section{Conclusions}

We presented an implementation of the RHMC algorithm for Wilson fermions in a
generic representation of the gauge group $\SU(N)$. We developed a bespoke code
-- HiRep -- and we used it to obtain some preliminary results for the $\SU(2)$
theory with two Dirac fermions in the adjoint representation, on a $16 \times
8^3$ lattice at fixed $\beta = 2.0$. Since we know very few about this theory,
we chose to proceed with caution in interpreting the results. We pointed out
that some results seem to fit in the standard scenario of a QCD-like theory (the
linear extrapolation of the PCAC mass with $1/\kappa$, the linear extrapolation
of the PS decay constant and of the vectorial meson mass with the PCAC mass),
some other results need more investigation to be understood (the behaviour of
the ratio $a M_{PS}^2/m$), other results seem to indicate non-standard phenomena
(the ratio $M_V/M_{PS}$ or the inversion of the axial and vectorial meson
masses). Finally we stressed the importance of checking the stability of these
results by increasing the volume and reducing the lattice spacing, especially in
a region close to the chiral limit.


\begin{thebibliography}{99}

\bibitem{Weinberg:1975gm}
S.~Weinberg, {\it Implications of dynamical symmetry breaking},  {\em Phys.
  Rev.} {\bf D13} (1976) 974--996.

\bibitem{Susskind:1978ms}
L.~Susskind, {\it Dynamics of spontaneous symmetry breaking in the {Weinberg-
  Salam} theory},  {\em Phys. Rev.} {\bf D20} (1979) 2619--2625.

\bibitem{Hill:2002ap}
C.~T. Hill and E.~H. Simmons, {\it Strong dynamics and electroweak symmetry
  breaking},  {\em Phys. Rept.} {\bf 381} (2003) 235--402 [{{\tt
  hep-ph/0203079}}].

\bibitem{Sannino:2008ha}
F.~Sannino, {\it Dynamical stabilization of the fermi scale: Phase diagram of
  strongly coupled theories for (minimal) walking technicolor and unparticles}
   [{{\tt 0804.0182 [hep-ph]}}].

\bibitem{corrigan:1979xf}
E.~Corrigan and P.~Ramond, {\it A note on the quark content of large color
  groups},  {\em Phys. Lett.} {\bf B87} (1979) 73.

\bibitem{Armoni:2003gp}
A.~Armoni, M.~Shifman and G.~Veneziano, {\it Exact results in
  non-supersymmetric large {N} orientifold field theories},  {\em Nucl. Phys.}
  {\bf B667} (2003) 170--182 [{{\tt hep-th/0302163}}].

\bibitem{Banks:1981nn}
T.~Banks and A.~Zaks, {\it On the phase structure of vector-like gauge theories
  with massless fermions},  {\em Nucl. Phys.} {\bf B196} (1982) 189.

\bibitem{Dietrich:2006cm}
D.~D. Dietrich and F.~Sannino, {\it Conformal window of {SU(N)} gauge theories
  with fermions in higher dimensional representations},  {\em Phys. Rev.} {\bf
  D75} (2007) 085018 [{{\tt hep-ph/0611341}}].

\bibitem{Foadi:2007ue}
R.~Foadi, M.~T. Frandsen, T.~A. Ryttov and F.~Sannino, {\it Minimal walking
  technicolor: Set up for collider physics},  {\em Phys. Rev.} {\bf D76} (2007)
  055005 [{{\tt 0706.1696 [hep-ph]}}].

\bibitem{Catterall:2007yx}
S.~Catterall and F.~Sannino, {\it Minimal walking on the lattice},  {\em Phys.
  Rev.} {\bf D76} (2007) 034504 [{{\tt 0705.1664 [hep-lat]}}].

\bibitem{Appelquist:2007hu}
T.~Appelquist, G.~T. Fleming and E.~T. Neil, {\it Lattice study of the
  conformal window in {QCD}-like theories} [{{\tt 0712.0609 [hep-ph]}}].

\bibitem{Shamir:2008pb}
Y.~Shamir, B.~Svetitsky and T.~DeGrand, {\it {Zero of the discrete beta
  function in {SU(3)} lattice gauge theory with color sextet fermions}} [{{\tt
  0803.1707 [hep-lat]}}].

\bibitem{Deuzeman:2008sc}
A.~Deuzeman, M.~P. Lombardo and E.~Pallante, {\it The physics of eight
  flavours} [{{\tt 0804.2905 [hep-lat]}}].

\bibitem{Clark:2006fx}
M.~A. Clark and A.~D. Kennedy, {\it Accelerating dynamical fermion computations
  using the rational hybrid monte carlo {(RHMC)} algorithm with multiple
  pseudofermion fields},  {\em Phys. Rev. Lett.} {\bf 98} (2007) 051601 [{{\tt
  hep-lat/0608015}}].

\bibitem{deldebbio:2008zf}
L.~Del~Debbio, A.~Patella and C.~Pica, {\it Higher representations on the
  lattice: numerical simulations. {SU(2)} with adjoint fermions} [{{\tt
  0805.2058 [hep-lat]}}].

\bibitem{Takaishi:2005tz}
T.~Takaishi and P.~de~Forcrand, {\it {Testing and tuning new symplectic
  integrators for hybrid Monte Carlo algorithm in lattice QCD}},  {\em Phys.
  Rev.} {\bf E73} (2006) 036706 [{{\tt hep-lat/0505020}}].

\bibitem{Luscher:2005rx}
M.~Luscher, {\it Schwarz-preconditioned {HMC} algorithm for two-flavour lattice
  {QCD}},  {\em Comput. Phys. Commun.} {\bf 165} (2005) 199--220 [{{\tt
  hep-lat/0409106}}].

\bibitem{DelDebbio:2005qa}
L.~Del~Debbio, L.~Giusti, M.~Luscher, R.~Petronzio and N.~Tantalo, {\it
  Stability of lattice {QCD} simulations and the thermodynamic limit},  {\em
  JHEP} {\bf 02} (2006) 011 [{{\tt hep-lat/0512021}}].

\bibitem{DelDebbio:2008wb}
L.~Del~Debbio, M.~T. Frandsen, H.~Panagopoulos and F.~Sannino, {\it Higher
  representations on the lattice: perturbative studies} [{{\tt 0802.0891
  [hep-lat]}}].

\end{thebibliography}

\end{document}